\documentclass[article,author-numerical,showpacs]{revtex4-2}

\usepackage{amsmath,amssymb,bm,graphicx,float,color,hyperref}

\usepackage[final]{changes}
\begin{document}

\title{Emergence of sector and spiral patterns \\
from a two-species mutualistic cross-feeding model}%

\author{Jiaqi.Lin, Hui. Sun, JiaJia Dong$^*$}
\affiliation{Department of Computer Science, Bucknell University}
\affiliation{Department of Mathematics, Cal State, Long Beach}
\affiliation{Department of Physics \& Astronomy, Bucknell University}
\email{jiajia.dong@bucknell.edu}

\begin{abstract}
The ubiquitous existence of microbial communities marks the importance of understanding how species interact within the community to coexist and their spatial organization. We study a two-species mutualistic cross-feeding model through a stochastic cellular automaton on a square lattice using kinetic Monte Carlo simulation. Our model encapsulates the essential dynamic processes such as cell growth, and nutrient excretion, diffusion and uptake. Focusing on the interplay among nutrient diffusion and individual cell division, we discover three general classes of colony morphology: co-existing sectors, co-existing spirals, and engulfment. When the cross-feeding nutrient is widely available, either through high excretion or fast diffusion, a stable circular colony with alternating species sector emerges. When the consumer cells rely on being spatially close to the producers, we observe a stable spiral. We also see one species being engulfed by the other when species interfaces merge due to stochastic fluctuation. By tuning the diffusion rate and the growth rate, we are able to gain quantitative insights into the structures of the sectors and the spirals.
\end{abstract}
\maketitle

\section{Introduction}

Understanding the emergence of complex patterns from simple dynamical rules and unraveling the rich behaviors is a continuous pursuit in physics. Microbial communities exist ubiquitously in nature, where multi-species occupy spatial spaces and interact with one another. Within microbial communities, how they survive,  expand,  and maintain species diversity provides much fodder for theoretical and experimental explorations. To make robust progress towards understanding a complex colony, it is important to identify the essential components within the colony \cite{konopka2009microbial}, the spatial structure of the colony \cite{davey2000microbial}, and the stability of the colony diversity \cite{tilman1999ecological} with various interactions within species and between species and the environment.

One way through which species interact with one another is by exchanging metabolites, namely metabolite cross-feeding. In a simple two-species system, there are several possible cross-feeding mechanisms depending on whether one or both species produce cross-feeding metabolite, and whether one or both species benefit from the process. Among all possible species interactions,  commensalism, syntrophy and mutualism \cite{stams2009electron,morris2013microbial,blanchard2015bacterial,smith2019classification} are often studied. In a system of two species, commensalism is defined where one species, the ``producer", produces metabolic byproduct that is essential for the growth of the other species, the ``consumer". The metabolite has a neutral effect on the growth of the producer. A common example of commensal interaction is that typically there are billions of the bacterium {\it Staphylococcus epidermidis} feeding on dead human skin cells while we, the ``producer'', do not react to them \cite{ghosh2013appraisal}. Syntrophy is similar to commensalism in that the producer excretes metabolites that benefit the consumer. However, the accumulation of the metabolite impacts the producer negatively by, for instance, changing the environment pH or access to oxygen. Thus the presence of the consumer helps reduce the environmental stress and thus the survival of the producer. Syntrophic interactions are closely associated with microbes under anoxic conditions and energy constraints \cite{morris2013microbial}.
Mutualism is when each species excretes a distinct metabolite that is essential for the other species. The accumulation of each metabolite can impact the producer neutrally or negatively, further underscoring the significance of the presence of consumer. Since population range expansion tends to create local homogeneity due to the ``founder effect'' \cite{excoffier2009genetic} while mutualistic interactions promotes species intermixing (see e.g. \cite{momeni2013strong}), it is of particular interest to us to investigate the interplay between cross-feeding and spatial structure of the colony.

In this study, we focus on a two-species \added{symmetric} mutualistic cross-feeding model in two-dimensional space, where one species excrete metabolic nutrients, to be taken up by the other species, and vice versa. While both species divide and grow, they also compete for space to expand.
Therefore, the colony spatial pattern depends crucially on the availability of the cross-feeding metabolites, which diffuse in open environments, and the local cross-feeding dynamics. We focus on the condition under which we can find stable coexistence of \deleted{of} both species and the corresponding colonial spatial structure. The stability of microbial communities \cite{tilman1999ecological} has many important implications, including the production of pharmaceutical products \cite{dao2018microbial}, indication of environmental condition \cite{tobor2005functional}, reflection on climate changes, and so on. We expect that through a simple growth model, we can begin to unravel the complex behaviors of ecological systems. 

Taking a similar approach as discrete cellular automata systems \cite{wolfram1983statistical,wolfram1984cellular,ermentrout1993cellular}, we devise a stochastic cellular automaton on a square lattice as a minimal model to investigate a symmetric mutualistic cross-feeding between two species. 
The dynamical evolution of the system is simulated using the kinetic Monte Carlo method (KMC)\cite{voter2007introduction,gillespie1976,gillespie1977}. This method is effective in describing variety of phenomena \cite{andersen2019practical}, including reaction-diffusion systems \cite{donev2010first,katsoulakis2003coarse}, structures and properties of materials \cite{piana2006three}, and equilibrium and non-equilibrium chemistry \cite{reuter2006first}. We use KMC to analyze a range of parameters that give rise to distinct colony morphology.
 
In Section \ref{sec:model}, we provide details of our model and simulation algorithm. We characterize the distinct colony patterns in Section \ref{sec:result} and analyze the interplay between cell growth and cross-feeding. Finally, we discuss in Section \ref{sec:sum} the generality of our findings. We suggest that the approach presented in the article provides preliminary predictions and powerful guidance on further studies, both in theory and in experiments, on multi-species colony pattern formation.

\section{Model Specification\label{sec:model}}

In this study, we focus on a symmetric set of mutualistic cross-feeding mechanism: Either species (type 1 or 2) has an individual growth rate $\lambda_{1,2}$ which depends on the availability of the metabolite, molecule $B$ (or $A$), produced by the other species and some other generic nutrient that is abundant in the environment. The metabolite excretion rates are $\gamma_{A,B}$ and the excreted molecule, if not taken up, diffuses with a diffusion coefficient of $D_{A,B}$. The yield $Y_{A,B}$ is a parameter that converts the nutrient molecules taken up by a species into cell biomass, typically measured as mass of nutrient per dry mass of cell or concentration of nutrient per unit optical density \cite{neidhardt1990physiology}. The dynamics of the system can be described in the following set of equations:

\begin{align}
\frac{\partial {\rho}_1}{\partial t} &= \lambda_1 (n_A)\rho_1;~~ \frac{\partial {\rho}_2}{\partial t} = \lambda_2 (n_B)\rho_2;\\
\frac{\partial n_A}{\partial t} &=\gamma_A \rho_2- \lambda_1 (n_A)\rho_1/Y_A+ D_A\cdot \nabla^2 n_A; \label{eq:nutrient1}\\
\frac{\partial n_B}{\partial t} &=\gamma_B \rho_1- \lambda_2 (n_B)\rho_2/Y_B+ D_B\cdot \nabla^2 n_B.  \label{eq:nutrient2}
\end{align}

\begin{figure}[!h]
\includegraphics[width=.5\textwidth]{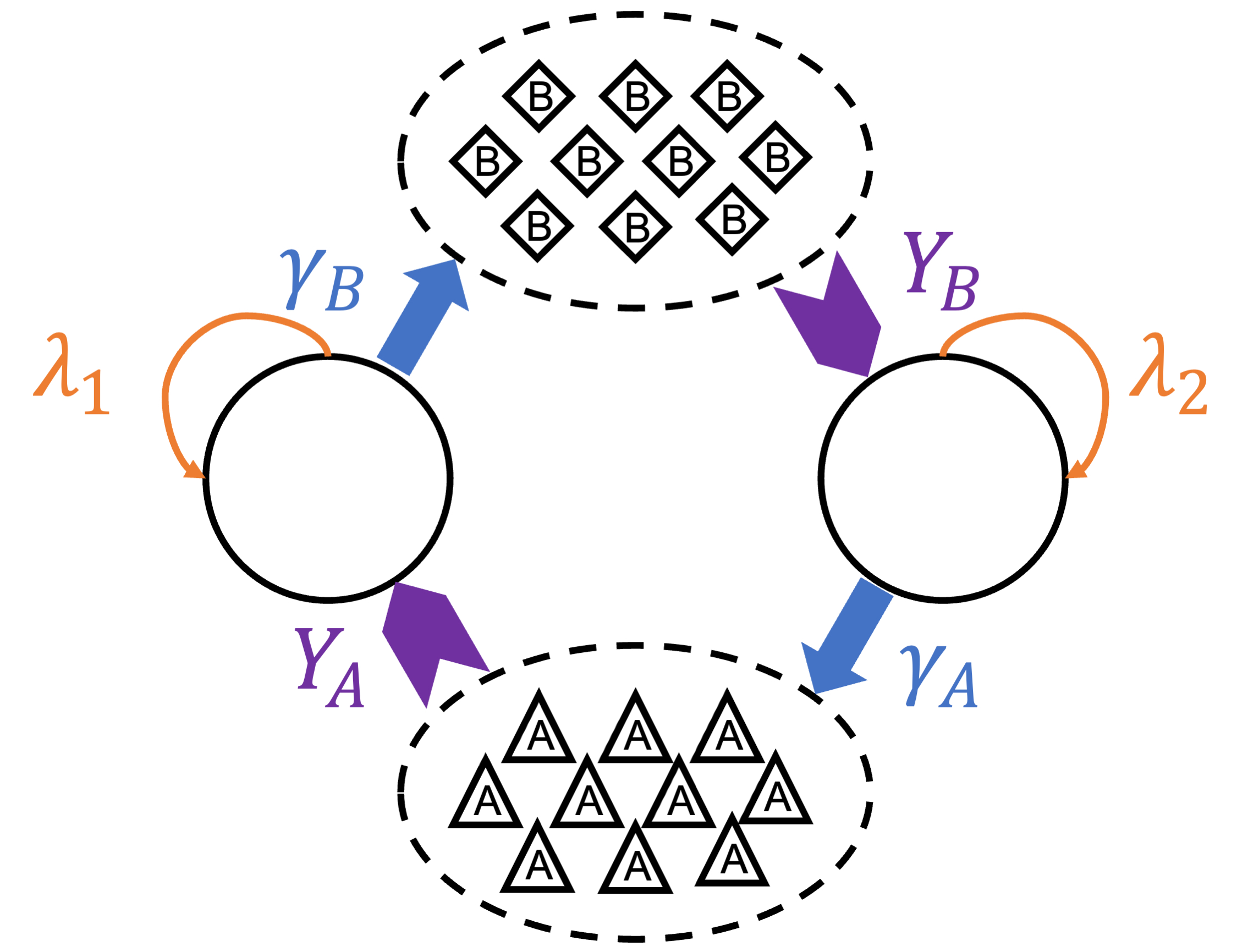}
\caption{{\bf Mutualistic cross-feeding scheme.}
As species 1 cells metabolize nutrient molecules $A$ (produced by species 2) at a growth rate $\lambda_1$, they excrete molecules $B$ with rate $\gamma_B$ as a metabolic by-product. Molecule $B$ is the essential nutrient for species 2 cells' growth and is taken up with yield $Y_B$. The same for the growth kinetics for cells of species 2.}
   \label{fig:schematic}
\end{figure}

Note that the cell concentration $\rho_{1,2}$ and the nutrient concentration $n_{A,B}$ depend on both space and time. The parameters and reactions of our model are summarized in a schematic shown in Fig.\ref{fig:schematic}.

The overarching question in this study is how spatiotemporal dynamics governed by cell growth and cross-feeding gives rise to distinct colony morphology and species diversity. We construct our simulation model on a 2-dimensional (2D) square lattice and incorporate cell doubling (growth), nutrient excretion and uptake, and nutrient diffusion. Compared to the nutrient molecules, the cell diffusion is much slower and is not included in this model. In this system, cells of both species are identical in size, occupying a single lattice site \deleted{of unit size} and cannot overlap. \added{Each lattice size is the same length of the cell.} Nutrient molecules have negligible spatial content and thus infinitely many of them can occupy the same lattice site. When a cell at lattice site $\vec{r}(x,y)$ is chosen to divide, the individual growth rate $\lambda_{1,2}$ is determined by $n_{A,B}$, the total nutrient concentration at $\vec{r}$ and its four nearest-neighbor (n.n.) lattice sites:
\begin{equation}
\lambda_{1,2}=\lambda_{1,2}^\ast\dfrac{\sum_{\rm{n.n.}}n_{A,B}}{\sum_{\rm{n.n.}}n_{A,B}+K_{A,B}}\label{eq:gr}
\end{equation}

Here $\lambda_{1,2}^\ast$ indicates the maximal single cell growth rate. The instantaneous growth rate is modified by a Monod factor \added{$K_{A,B}$} \cite{monod1949growth} to reflect the nutrient dependence, as shown in Eq.\ref{eq:gr}. The daughter cell is placed in one of the four n.n. lattice sites, chosen at random provided that it is empty. This is similar to the Eden model in surface growth \cite{eden1961two, barabasi1995fractal}. \deleted{When}\added{If} a nutrient excretion process is chosen, then \added{a randomly chosen producer cell produces} \deleted{the}\added{a} \deleted{newly produced} nutrient molecule \added{which} is placed at one of the n.n. sites of the producer cell. \added{If an uptake process is chosen to occur,} \deleted{When} a cell \added{is picked at random.} \deleted{takes up a nutrient molecule,} We compute the nutrient concentration in the 4 n.n. sites and the cell picks up one nutrient molecule chosen at random. When nutrient molecules diffuse, each diffusion step puts them in one of the four n.n. sites as well. In our simulation scheme, cells that do not divide due to spatial limitation still consumes nutrients and excrete metabolic by-product in an attempt to mimic a basal level of metabolic maintenance. This, however, can be revised into other scenarios to capture different metabolic processes which we will not delve into here.

The system is typically initialized with an inoculum patch of $L\times L$ lattice sites with $\rho_0 L^2$ cells, half of each species randomly distributed within the patch. $L^2$ is much smaller than the final colony size and $\rho_0$ is 1/4 unless otherwise specified. We implement the kinetic Monte Carlo algorithm\cite{gillespie1976,gillespie1977} using the direct method to simulate the dynamics of the system outlined in Fig.\ref{fig:schematic}. The individual rates for each process are summarized in Table \ref{tab:rxn}. Specifically, $\rho_{1,2}$ is the total number of cell type $1$ (or $2$), while $n_{A,B}$ is the total number of nutrient molecule $A$ (or $B$) in the entire system at a given time $t$. The propensity $k_j$ of each reaction, which scales with reaction probability, is determined by the respective reaction rate (time-independent) and the amount of reagents in the system at time $t$, and $k_0$ is the sum of all propensities.

\begin{table}[t]
   
\begin{center}
  \begin{tabular}{ | c| l |l|l|}
  \hline
  \#&reaction&propensity&notes\\
    \hline\hline
  1& type 1 cell divide & $k_1 = \lambda_1 \rho_1$& $\lambda_1 = \lambda_1^\ast\dfrac{n_A}{n_A+K_A} $; \\
  2& type 1 cell excrete $n_B$& $k_2 = \gamma_B \rho_1$&\\
  3& type 1 cell take up $n_A$& $k_3=\dfrac{\lambda_1}{Y_A}\rho_1$&\\ 
  4&type 2 cell divide & $k_4 = \lambda_2 \rho_2$& $\lambda_2= \lambda_2^\ast \dfrac{n_B}{n_B+K_B}$; \\
  5& type 2 cell excrete $n_A$& $k_5= \gamma_A \rho_2$&\\  
6&  type 2 cell take up $n_B$& $k_6= \dfrac{\lambda_2}{Y_B}\rho_2$&\\
  7& nutr. $B$ diffuses & $k_7 =  D_B n_B$& \\
8& nutr. $A$ diffuses & $k_8 =  D_A n_A$ &\\
  \hline
 && $k_0=\sum_{j=1}^8 k_j$ &\\
 \hline
  \end{tabular}
\end{center}
 \caption{Processes involved in the system and the corresponding rates.}
    \label{tab:rxn}
\end{table}

Let ${\rm r}_1, {\rm r}_2 \in (0,1)$ be two independent, uniformly distributed random numbers. At time $t$, the $j$-th reaction ($j\in [1,8]$) takes place if:

\[\sum_{i=1}^{j-1}k_i < k_0\cdot{\rm r}_1\leq\sum_{i=1}^{j}k_i. \]
Afterwards $t$ is advanced by $\Delta t$ where
\[\Delta t =-\dfrac{\ln{\rm r}_2}{k_0}.\]

The iteration continues until the system reaches a final time $T_{\rm max}$, typically after an obvious pattern is established. \added{The system is large enough so that the colony can keep growing without running into the boundaries.} When we refer to growth rate in the remainder of the study, we almost exclusively mean the {\it maximal} single cell growth rate. For ease of notation we will therefore drop the asterisk and denote it as $\lambda_{1,2}$. Throughout this study, the parameters are symmetric for both species, namely they will have the same individual growth rate, nutrient excretion and uptake rates, as well as the nutrient diffusion rate. Each species produce one type of nutrient and are simultaneously consuming the nutrient produced by the other species.

\section{Results\label{sec:result}}

\subsection{Sectors with high nutrient diffusion\label{sec:sector}}

When the essential nutrients for both species' growth are readily available, it has been shown that an initially mixed populations of two species will segregate into sector-like domains as a result of random fluctuations at the expanding colony frontier \added{both experimentally \cite{Hallatschek19926,muller2014genetic} and theoretically \cite{korolev2011competition,lavrentovich2014asymmetric,menon2015public}}. The domain boundaries perform super-diffusive random walks with respect to the radial growth. In the presence of mutualistic cross-feeding, when sufficient nutrients are produced by the cells, the system should behave in the same way as the nutrient-rich scenario in the absence of cross-feeding. We set out to first verify our model in this limit by comparing the following cases: 1) the environment provides nutrients for both species without cross-feeding; 2) cross-feeding with high nutrient excretion rates $\gamma_{A,B}$ and high diffusion coefficients $D_{A,B}$. 

\begin{figure}[!h]
\includegraphics[width=.5\textwidth]{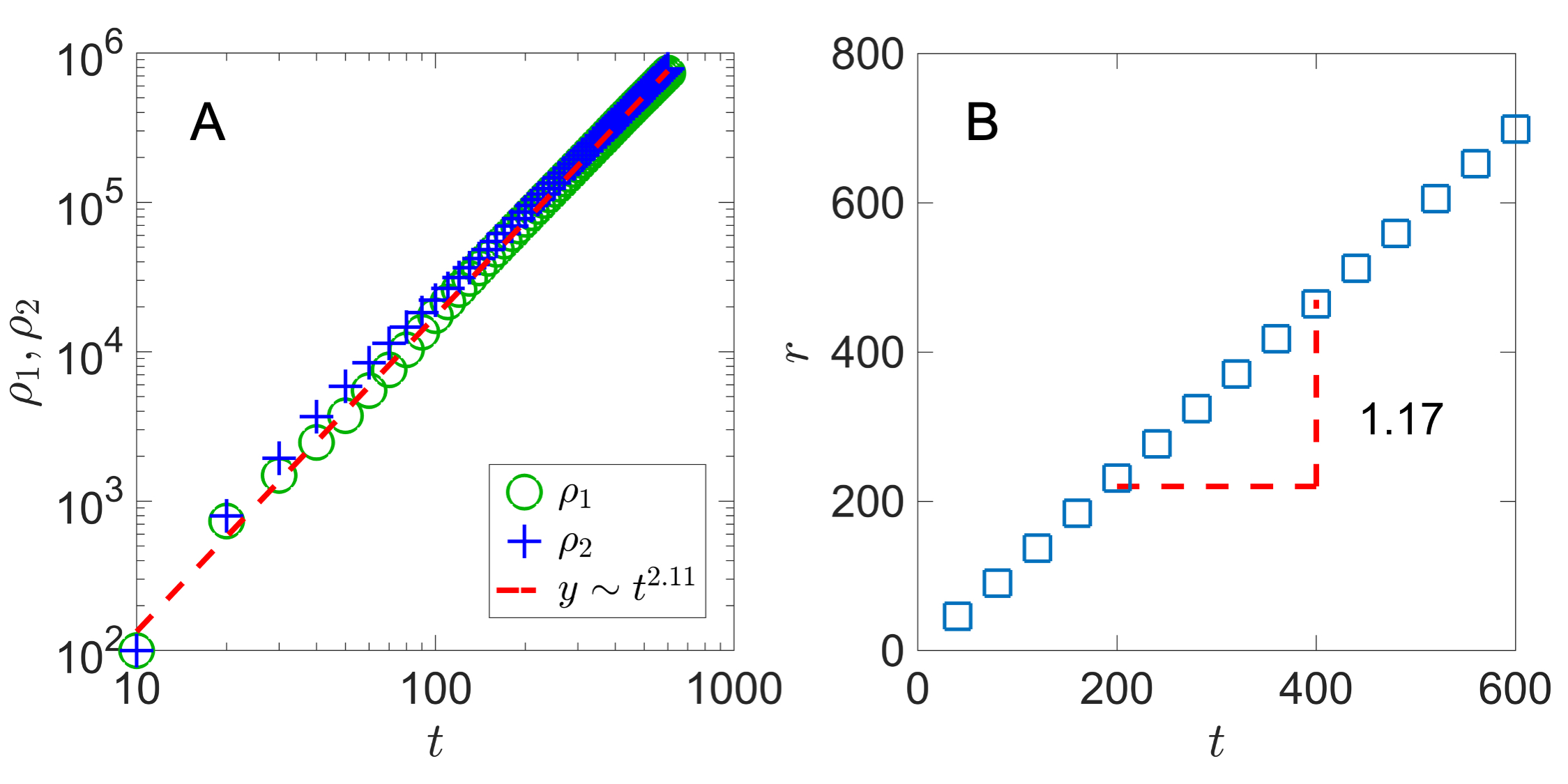}
    \caption{{\bf Colony population and radius with pre-seeded nutrients.} A) Colony population $\rho_{1,2}(t)$ and B) colony radius $r(t)$. Equal number of species 1 and 2 are seeded in a $20\times20$ patch with initial density $\rho_0 =1/4$. $\lambda_{1,2} = 1, D_{A,B} = 1, Y_{A,B} = 1$,  with nutrients placed throughout the lattice.}
        \label{fig:sector_pop}
\end{figure}

In the first case, the 2D system is initially seeded with an equal amount of species 1 and 2 cells, 50 of each type randomly distributed in a $20\times20$ lattice. Every lattice site, namely the ``environment'', starts with 15 of $A$ and 15 of $B$ molecules. Since nutrients are widely available, both species grow without relying on cross-feeding. However, they are subject to the spatial constraint and only cells adjacent to at least one empty lattice site can divide. Fig.\ref{fig:sector_pop} shows the population of each species $\rho_{1,2}(t)$ over time. In the early stage ($t\lesssim 20$), the colony quickly expands with little spatial constraint. It then transitions to a constant radial expansion. In our system, only cells at the colony frontier have access to empty space to grow and thus contributing to colony expansion. Assuming an approximate circular colony of radius $r$, we have the total colony size $\rho \approx \pi r^2$ and the number of cells at the expanding frontier $2 \pi r$. The radial expansion of the colony is approximately: 
\begin{equation}
    \dfrac{d}{dt}\left(\pi r^2\right) = \lambda \cdot 2\pi r \rightarrow \dfrac{dr}{dt} = \lambda, \rho \sim~t^2\label{eq:exp}
\end{equation}
This gives a constant colony expansion speed that depends on the individual growth rate $\lambda$ and the total population within the colony grows quadratic with time.

In Fig.\ref{fig:sector_pop}, we show the population $\rho_{1,2}(t)$ and the colony radius $r(t)$ for a system with nutrients available from the environment. In our simulation, we determine the colony radius $r(t)$ by measuring the distances between the center and all cells adjacent to an empty lattice site, and then taking the average among all the distances. The colony expands radially with both cell populations grow at $t^{2.11}$. The expansion speed $dr/dt$, measured to be 1.17, is consistent with the individual growth rate $\lambda_{1,2} = 1.$

\begin{figure}[!h]
\includegraphics[width=.5\textwidth]{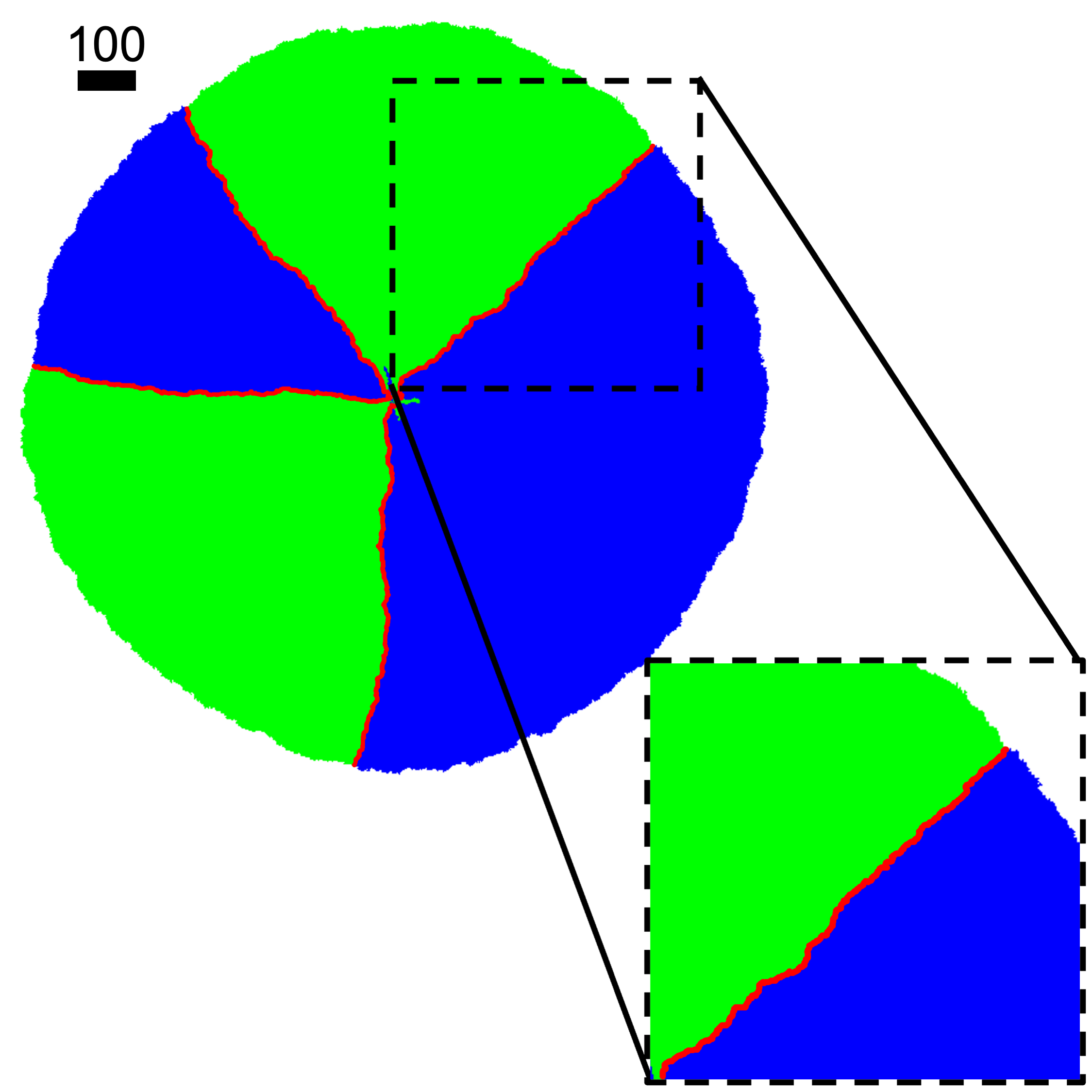}
    \caption{{\bf Colony morphology with pre-seeded nutrients.} Colony snapshot at $t= 600$. Equal number of species 1 and 2 are seeded in a $20\times20$ patch with initial density $\rho_0 =1/4$. $\lambda_{1,2} = 1, D_{A,B} = 1, Y_{A,B} = 1$ with nutrients placed throughout the lattice. Scale bar indicates the width of 100 cells.}
    \label{fig:sector}
\end{figure}

The two species within the colony are separated by meandering interfaces. A colony snapshot at $t= 600$ with interface marked in red is shown in Fig.\ref{fig:sector}. As expected, the initially mixed two populations segregate into sectors as the individual cells divide and the colony expands. Because the fluctuation of the interfaces is intimately related to the colony morphology, we analyze the inter-species interfaces to see whether the fluctuations resemble a standard random walk. We adopt a similar quantification method as discussed in Ref.\cite{Hallatschek19926} with details included in \deleted{the Appendix}\added{S1}.

\begin{figure}[!h]
\includegraphics[width=.5\textwidth]{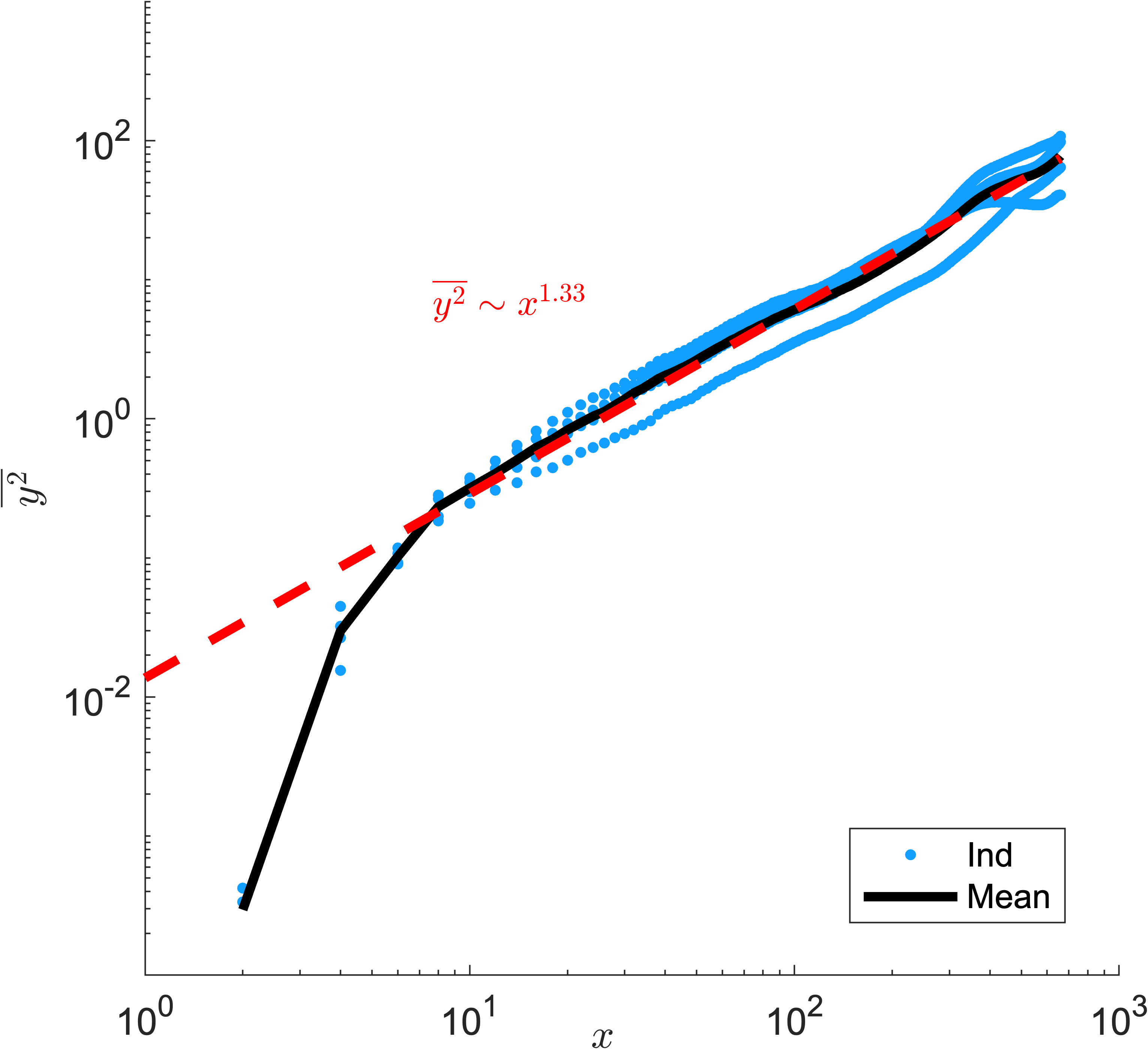}
    \caption{{\bf Species interface fluctuation in sectors.} Interface mean square displacement as a function of the sliding-window size $x$ with nutrients from the environment. Symbols (blue, color online): 4 individual interface trajectories in Fig.\ref{fig:sector}; Black line: average of the 4 trajectories. Red line: power-law fit with exponent $1.33\pm 0.18$. Equal number of species 1 and 2 are seeded in a $20\times20$ patch with initial density $\rho_0 =1/4$. $\lambda_{1,2} = 1, D_{A,B} = 1, Y_{A,B} = 1$ with nutrients placed throughout the lattice.}
    \label{fig:interface}
\end{figure}

 For an interface that fluctuates like a standard random walk, we expect the mean-square displacement to scale linearly with time, $\overline{y^2}\sim t$. In our case, $\overline{y^2}\sim t^{4/3}$ as shown in Fig.\ref{fig:interface} indicates a super-diffusive behavior which is consistent with the findings in Ref. \cite{Hallatschek19926}. In this scenario, the adjacent interfaces extends linearly with time while their fluctuation grows sub-linearly. This establishes a long-time stability of the sector morphology in 2-dimensional space.
 
Now we turn to the second case to see how nutrient excretion rate $\gamma_{A,B}$ under high nutrient diffusion $D_{A,B}$ impacts the colony morphology. Intuitively, when cross-feeding produces abundant nutrients through high excretion and/or high yield, while fast diffusion brings nutrients to the growing frontier, the colony dynamics should resemble the first case where nutrients are readily available from the environment and the two species do not depend on each other spatially. 

\begin{figure}[!h]
\includegraphics[width=.5\textwidth]{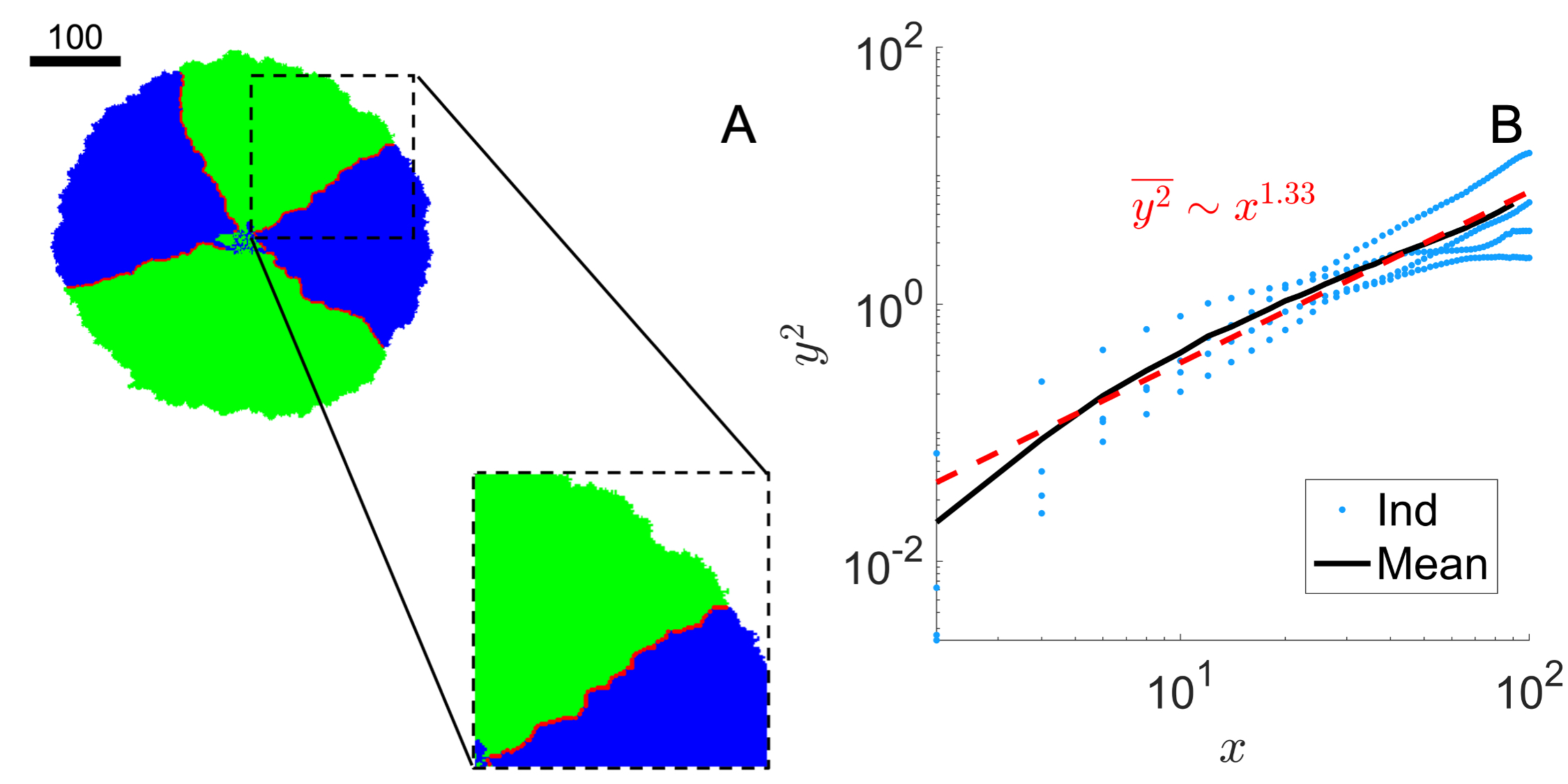}
\caption{{\bf Colony morphology and interface fluctuation with high nutrient excretion.} A) Colony snapshot at $t= 140$ for cross-feeding with high nutrient excretion. Inoculum condition is the same as in Fig.\ref{fig:sector}. $\lambda_{1,2} = 1, D_{A,B}= 500, \gamma_{A,B} = 10, Y_{A,B} = 1$.  Scale bar indicates the width of 100 cells. B) Interface mean square displacement as a function of the sliding-window size $x$ with high nutrient excretion. Symbols (blue, color online): 4 individual interface trajectories. Black line: average of the 4 individual ones. Red line: power-law fit with exponent $1.33\pm 0.40$. }
   \label{fig:high_exc}
\end{figure}

\begin{figure}[!h]
\includegraphics[width=.5\textwidth]{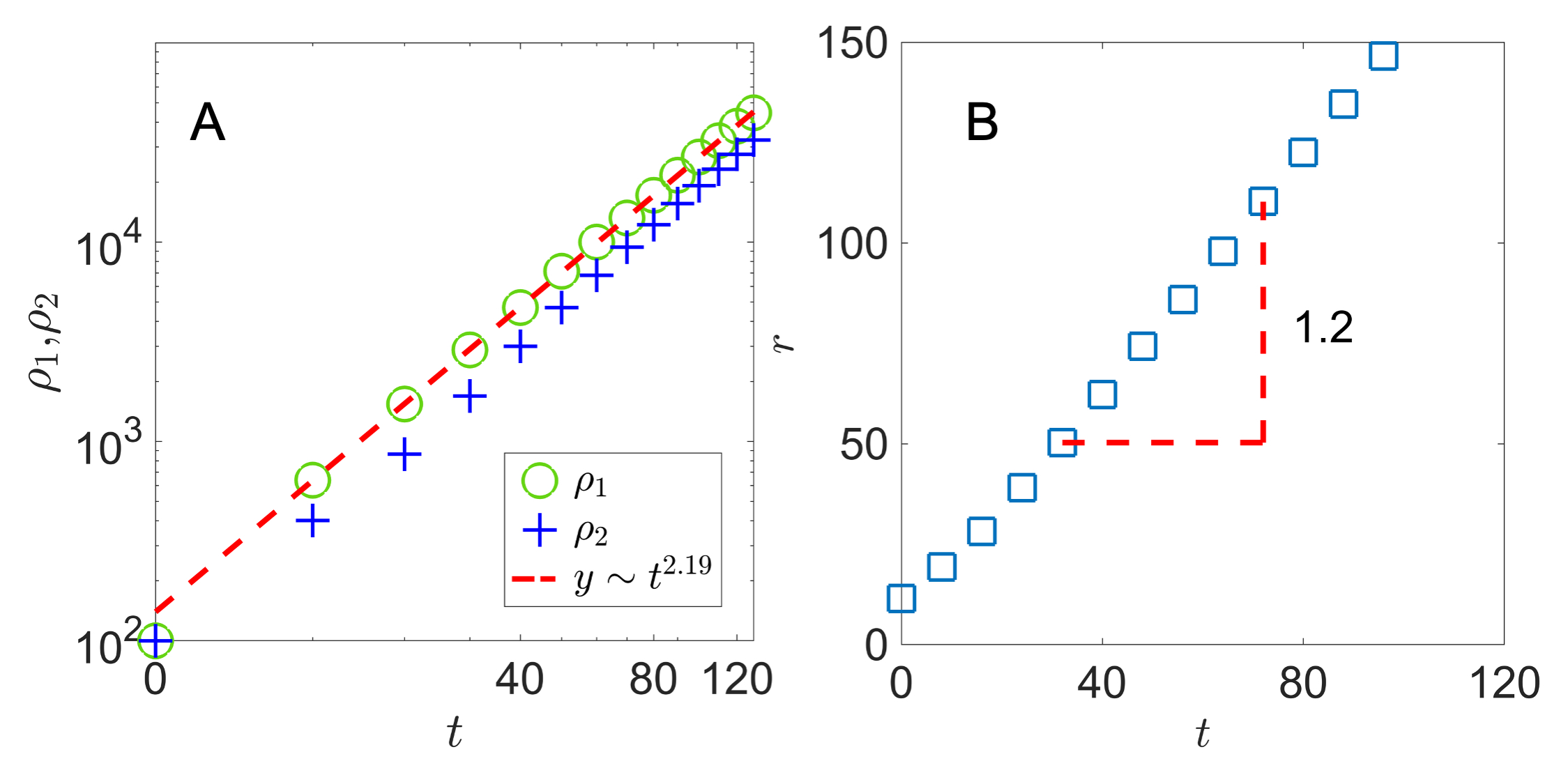}
\caption{{\bf Colony population and radius with high nutrient excretion.} A) Colony population $\rho_{1,2}(t)$ and  B) colony radius $r(t)$. Equal number of species 1 and 2 are seeded in a $20\times20$ patch with initial density $\rho_0 =1/4$. $\lambda_{1,2} = 1, D_{A,B} = 500, \gamma_{A,B} = 10, Y_{A,B} = 1$.
}
   \label{fig:high_exc_pop}
\end{figure}

When we set both excretion rates and diffusion to be high, we indeed see the colony pattern develops similarly as the case where nutrients are pre-seeded in the environment. In Fig.\ref{fig:high_exc}, we show a colony snapshot at $t=140$ with cross-feeding at a high nutrient excretion rate $\gamma_{A,B} = 10$ and high nutrient diffusion rate $D_{A,B}=500$. Similar to Fig.\ref{fig:sector}, the colony again separates into sectors of same-type species. The interface mean-square displacement $\overline{y^2}\sim t^{1.33}$, same as in Fig.\ref{fig:interface}. Fig.\ref{fig:high_exc_pop} shows the cell population $\rho_{1,2}(t)$ and colony radius $r(t)$. The overall population displays a quadratic growth while $r^2$ increases faster than $\rho$. This is a result of deviation from circular colony growth: We measure the colony circumference roughness as defined in \cite{barabasi1995fractal} for the two growth conditions shown in Figs.\ref{fig:sector} and \ref{fig:high_exc}, and indeed find an increase in roughness in the latter case.

\begin{figure}[!h]
\includegraphics[width=.5\textwidth]{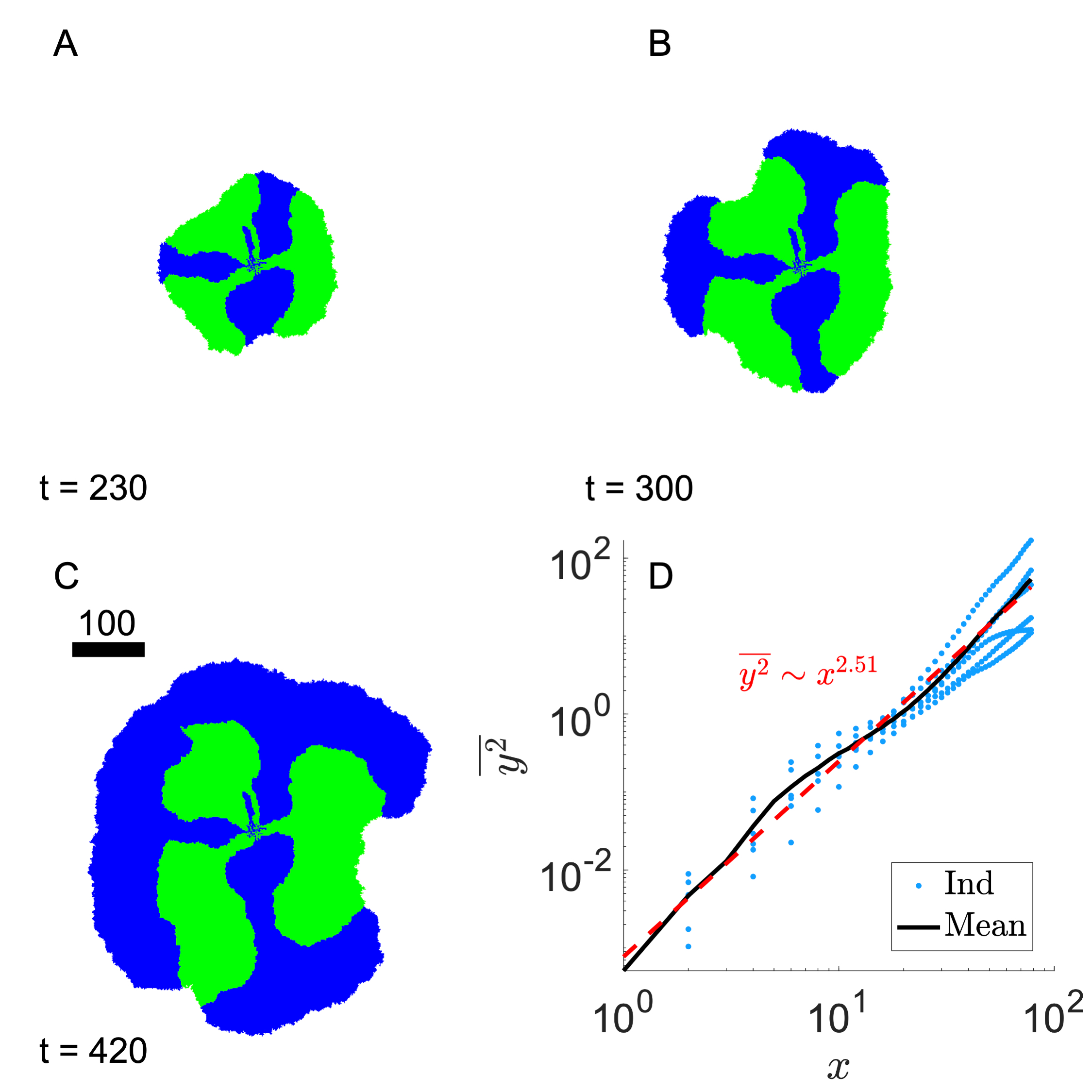}
\caption{{\bf Coalesce of species interfaces with high nutrient diffusion.} A)- C) Colony snapshots at $t= 230, 300$, and $420$. D) Interface mean square displacement as a function of the sliding-window size $x$. Symbols (blue, color online): 6 individual interface trajectories. Black line: average of the 4 individual ones. Red line: power-law fit with exponent $2.51$. Inoculum condition is the same as in Fig.\ref{fig:sector}. $\lambda_{1,2} = 1, D_{A,B} = 500, \gamma_{A,B} = 1, Y_{A,B} =1$. Scale bar indicates the width of 100 cells. }   \label{fig:high_diff}
\end{figure}

The species' reliance on cross-feeding develops as we dial down the nutrient excretion rate while holding nutrient diffusion constant. In Fig.\ref{fig:high_diff}, we keep $D_{A,B} = 500$ as in Fig.\ref{fig:high_exc} while reducing $\gamma_{A,B}$ to 1. The nutrient concentration at the expanding front is no longer keeping up with diffusion in the same radial direction. As a result, the interface fluctuation increases over time, $\overline{y^2}\sim r^2 \sim t^{2.5}$. This means at least some of the interfaces will collide and annihilate in a two-dimensional system, as seen in  Fig.\ref{fig:high_diff} at later times. 

The coalesce of interfaces clearly points to the possibility of one species being crowded out, or ``engulfed'', by the other when all of the interfaces eventually merge. \added{The specific time at which engulfment occurs will depend on the dynamics of the system. In this case, the colony expands radially with constant speed. However, the mean square displacement of the interface $\overline{y^2}$ grows faster than $t^2$, shown in Fig.\ref{fig:high_diff}(D). This means that as time increases, the two interfaces will eventually coalesce.}

 \deleted{However, if the mean displacement $\overline{y^2}$ grows faster than $t^2$, then engulfment will eventually happen in a 2-D system.  }

\subsection{Spirals emerge when cross-feeding limits growth}

In the previous section, we discuss the emergence of sector patterns under high nutrient availability and the merging of sectors when nutrient concentration in the radial direction drops below the consumption. The stochasticity at the growing tip of the interface gives rise to the sector patterns of the colony. Here we hone in on the case where globally there is a scarcity of nutrients due to low excretion and low diffusion, thus the proliferation of either species depends on the availability of the cross-feeding nutrients.

Starting the system with a square patch of side-length $20$ and initial density $\rho_0 = 1/4$ as before, we set the excretion rate $\gamma_{A,B}$ and the diffusion rate $D_{A,B}$ to be 1. Here with $\gamma_{A,B} = Y_{A,B} = 1$. This means each producer produces one molecule and the consumer needs one molecule to be at half-maximal growth rate.

\begin{figure}[!h]
\includegraphics[width=.5\textwidth]{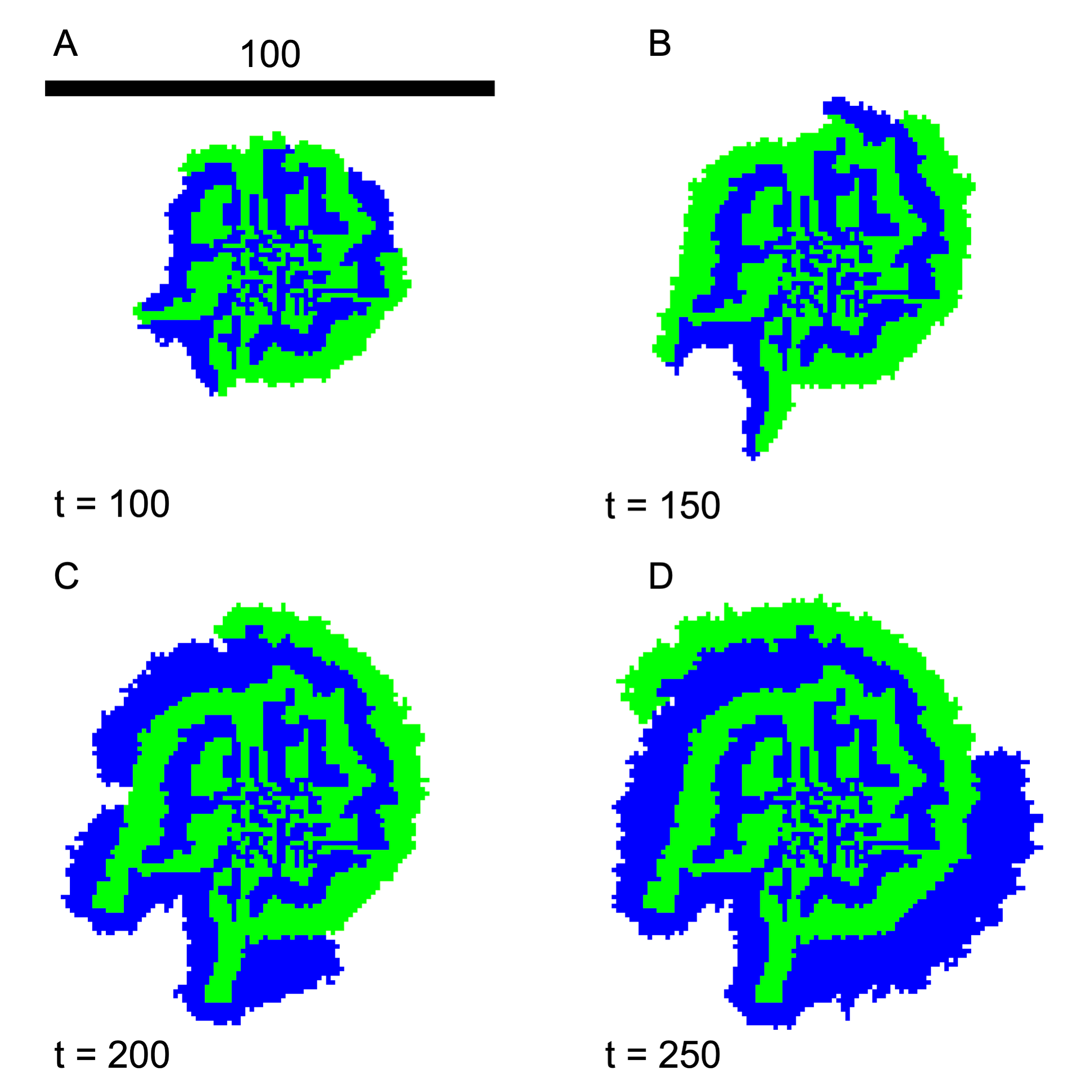}
\caption{{\bf Early emergence of a spiral pattern.} A)-D) Colony snapshots from $t=  100$ to $250$. Equal number of species 1 and 2 are seeded in a $20\times20$ patch with initial density $\rho_0 =1/4$. $\lambda_{1,2} = 1, D_{A,B} = 1, \gamma_{A,B} = 1, Y_{A,B} =1$. Scale bar indicates the width of 100 cells.}   \label{fig:colony_spiral}
\end{figure}

\begin{figure}[!h]
\includegraphics[width=.5\textwidth]{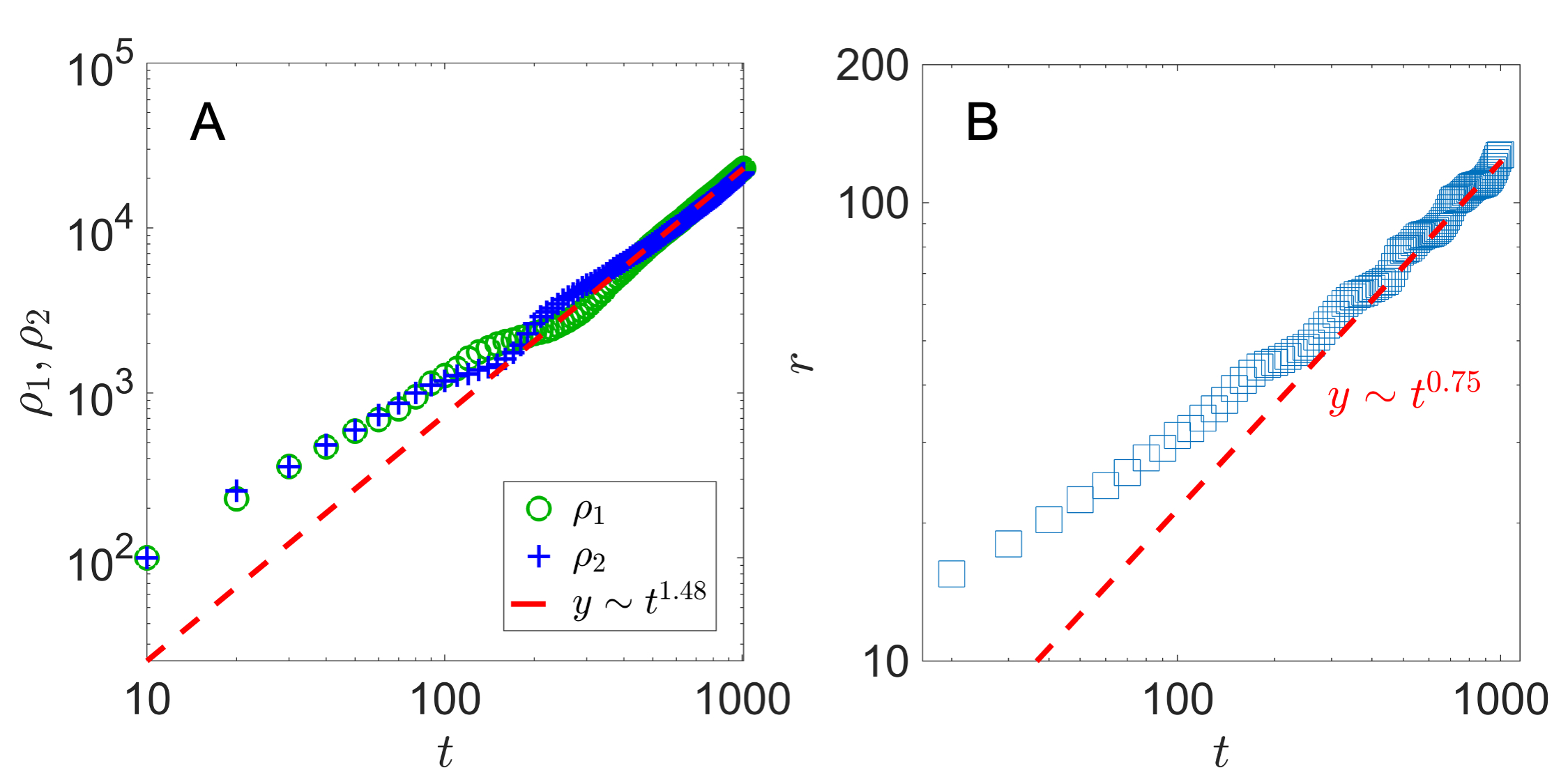}
    \caption{{\bf Colony population and radius in a spiral pattern.} A) Colony population $\rho_{1,2}(t)$ and B) colony radius $r(t)$ (right). Equal number of species 1 and 2 are seeded in a $20\times20$ patch with initial density $\rho_0 =1/4$. $\lambda_{1,2} = 1, D_{A,B} = 1, \gamma_{A,B}=1, Y_{A,B} = 1.$}
    \label{fig:colony_spiral_pop}
\end{figure}

We are surprised to see a stable spiral pattern of the colony emerge after a transient period. In Fig.\ref{fig:colony_spiral}, we show the nascent colony patterns at $t= 100, 150, 200$ and $250$. In the early stage of the colony growth ($t\lesssim  100$), the small domains of the two species quickly merge due to the short separation distance and the resulting interfaces start to bend from the radial direction. Since cell type 1 needs $n_A$ (produced by cell type 2) to grow, only the ones near a type 2 cell are growing when diffusion is slow. The same is true for cell type 2. Both types of cells grow along the interface between them, giving rise to multiple branches of spirals with cell growth in the tip of the branches and in the radial direction. Unlike the quadratic population growth ($\rho\sim t^2$) -- equivalently constant speed radial growth -- in the sector pattern case discussed in the previous section, the overall population in the spiral pattern case grows more slowly due to both nutrient and spatial limitation. In Fig.\ref{fig:colony_spiral_pop}, we see that both populations grow $\sim t^{3/2}$, and the colony radius $r$ expands sub-linearly in time, indicating a decreasing radial advancing speed. 

\begin{figure}[!h]
\includegraphics[width=.5\textwidth]{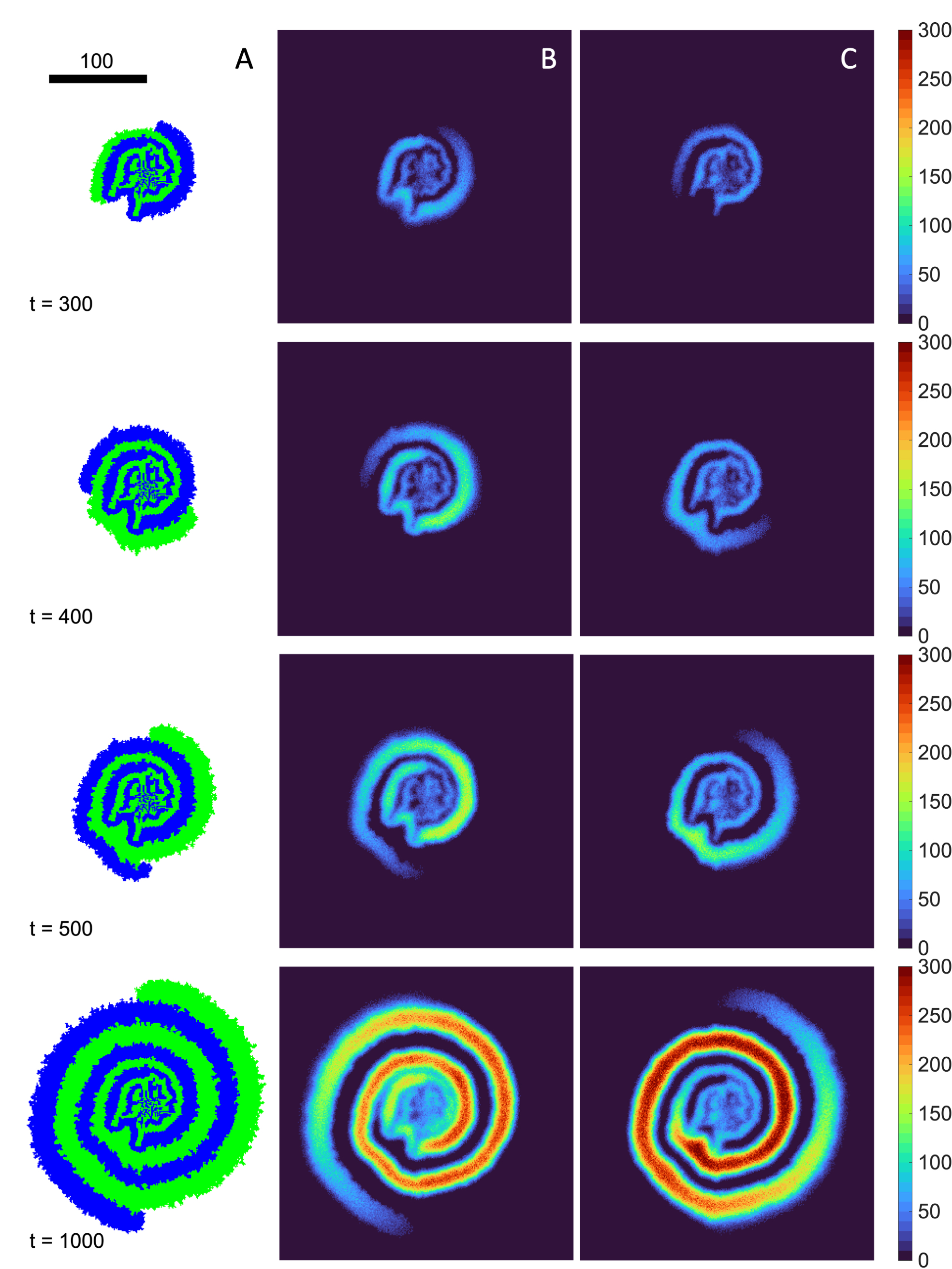}
\caption{{\bf Development of spiral pattern and the corresponding nutrient profiles. } A) Snapshots of the colony morphology at $t= 300$, $400$, $500$ and $1000$. B) and C) show the corresponding nutrient profiles. $\lambda_{1,2} = 1, D_{A,B} = 1, \gamma_{A,B} = 1, Y_{A,B} =1$. Scale bar indicates a width of 100 cells. }   \label{fig:spiral_nutrient}
\end{figure}

To further investigate the colony radial growth, we examine the nutrient profiles together with the colony morphology, shown in Fig.\ref{fig:spiral_nutrient}. We observe that, with the slow diffusion $D_{A,B} = 1$, the regions with metabolites closely mirror the spatial distribution of the producer cells. The nutrient depletion regions also mirror the spatial distribution of the consumers. In addition, the bounds of the colony does not exceed that of the nutrient profiles. Collectively, this indicates that the nutrient diffusion limits the colony expansion. The tip of the spiral has a low metabolite concentration because of the newly founded producers. The nutrient profiles shown in \deleted{the middle and right columns of} Fig.\ref{fig:spiral_nutrient}\added{B and C} are the net result of nutrient production and consumption. Within the cross-section of a branch of producers, the overall nutrient production per unit time is constant due to the fixed number of producers confined in space. With regard to consumption, the diffusion of nutrient determines how far the consumers can grow. As visualized in \deleted{the middle and right columns of the nutrient profiles in} Fig.\ref{fig:spiral_nutrient}\added{B and C}, there is a build-up of nutrient in the center of the cross-section while the concentration drops to zero in the radial directions due to complete consumption of nutrients that diffuse into the region. We will see next that this indeed impacts the colony morphology when individual growth rates are increased.

\begin{figure}[!h]
\includegraphics[width=.5\textwidth]{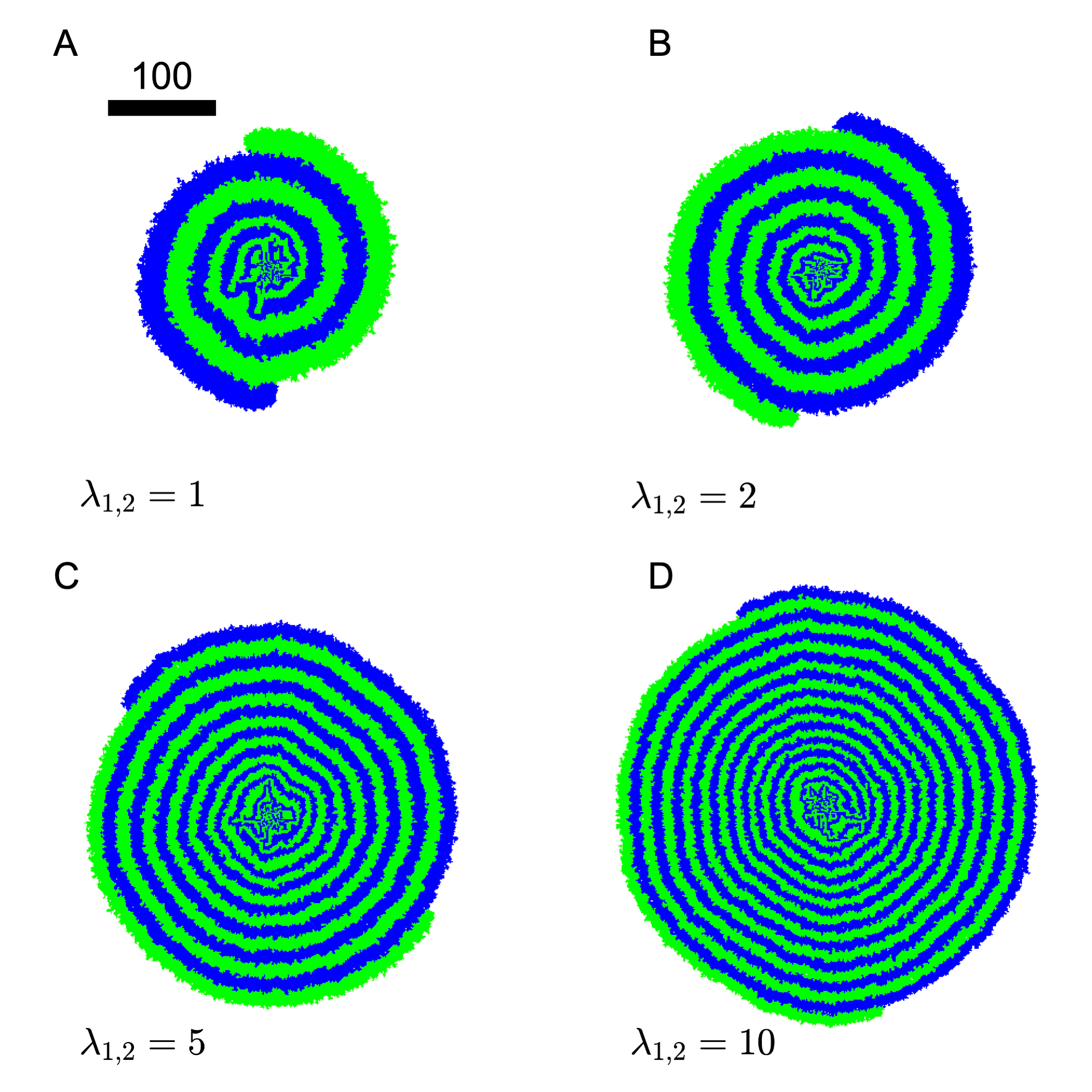}
\caption{{\bf Colony morphology with different individual cell growth rates.} Colony snapshots at $t=1000$ with $\lambda_{1,2} = $ A) $1$, B) $2$, C) $5$, D) $10$. In all cases, $D_{A,B} = 1, \gamma_{A,B}=1, Y_{A,B} = 1.$ Scale bar indicates a width of 100 cells.}
\label{fig:GR_spiral}
\end{figure}

\begin{figure}[!h]
\includegraphics[width=.5\textwidth]{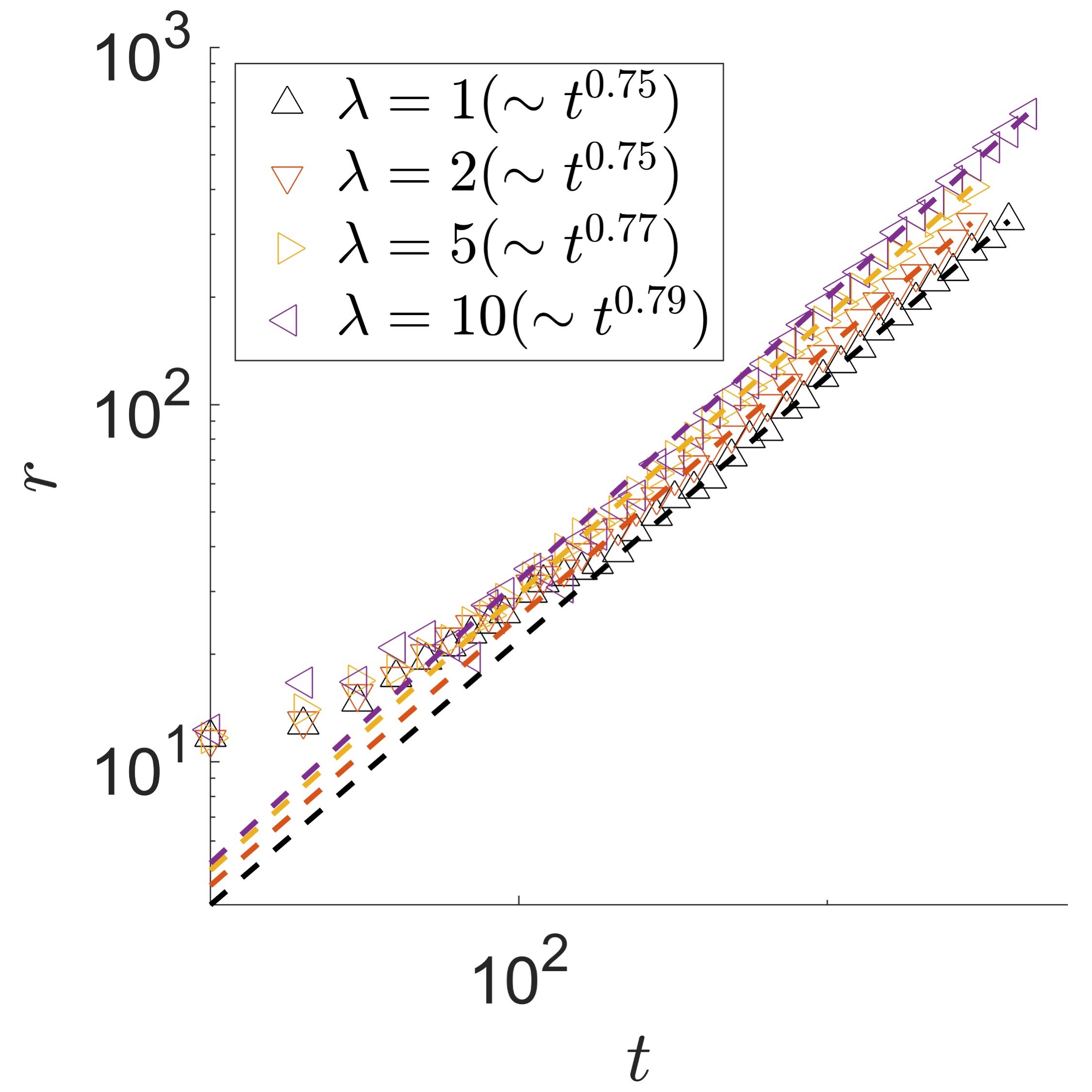}
    \caption{{\bf Radius $r(t)$ of a spiral pattern.} $\lambda_{1,2} = 1,2,5,10$. In all cases, $D_{A,B} = 1, \gamma_{A,B}=1, Y_{A,B} = 1.$ The power-law fit is performed to times after a clear spiral is established.}
\label{fig:radius}
\end{figure}

The characteristics and evolution of the spiral patterns depend on the single cell growth rate $\lambda_{1,2}$ when we keep the other parameters the same. In Fig.\ref{fig:GR_spiral}, snapshots of the colony are taken at $t=1000$ for growth rates $\lambda_{1,2} = 2, 5$ and $10$ along with $\lambda_{1,2} =1$ for comparison. With larger growth rates, the overall colony is larger in radius after growing for the same amount of time. In Fig.\ref{fig:radius}, we can see a slight growth rate dependence in the colony radial expansion speed $d r(t)/dt$, and in all cases the radial expansion is slowing down over time. Unlike the sector patterns discussed earlier, the radial expansion in the spiral patterns comes mainly from the branch wrapping around. We can see from Fig.\ref{fig:GR_spiral} that, with a higher growth rate, the colony has a ``thinner'' wrapping branch. As alluded to previously, the steady state of nutrient concentration is established within a spiral branch. Since consumers can only get metabolites through diffusion $D_{A,B}$ and the time scale of significance is the growth rate $\lambda_{1,2}$, the characteristic distance for the metabolite diffusion, and thus the characteristic width of the branch, can be estimated using dimensional analysis as $w(t) \sim \sqrt{2D_{A,B}/\lambda_{1,2}}$ at a given time $t$. Therefore, as $\lambda_{1,2}$ increases, we see a ``thinner'' wrapping branch emerging.

\begin{figure}[!h]
\includegraphics[width=.5\textwidth]{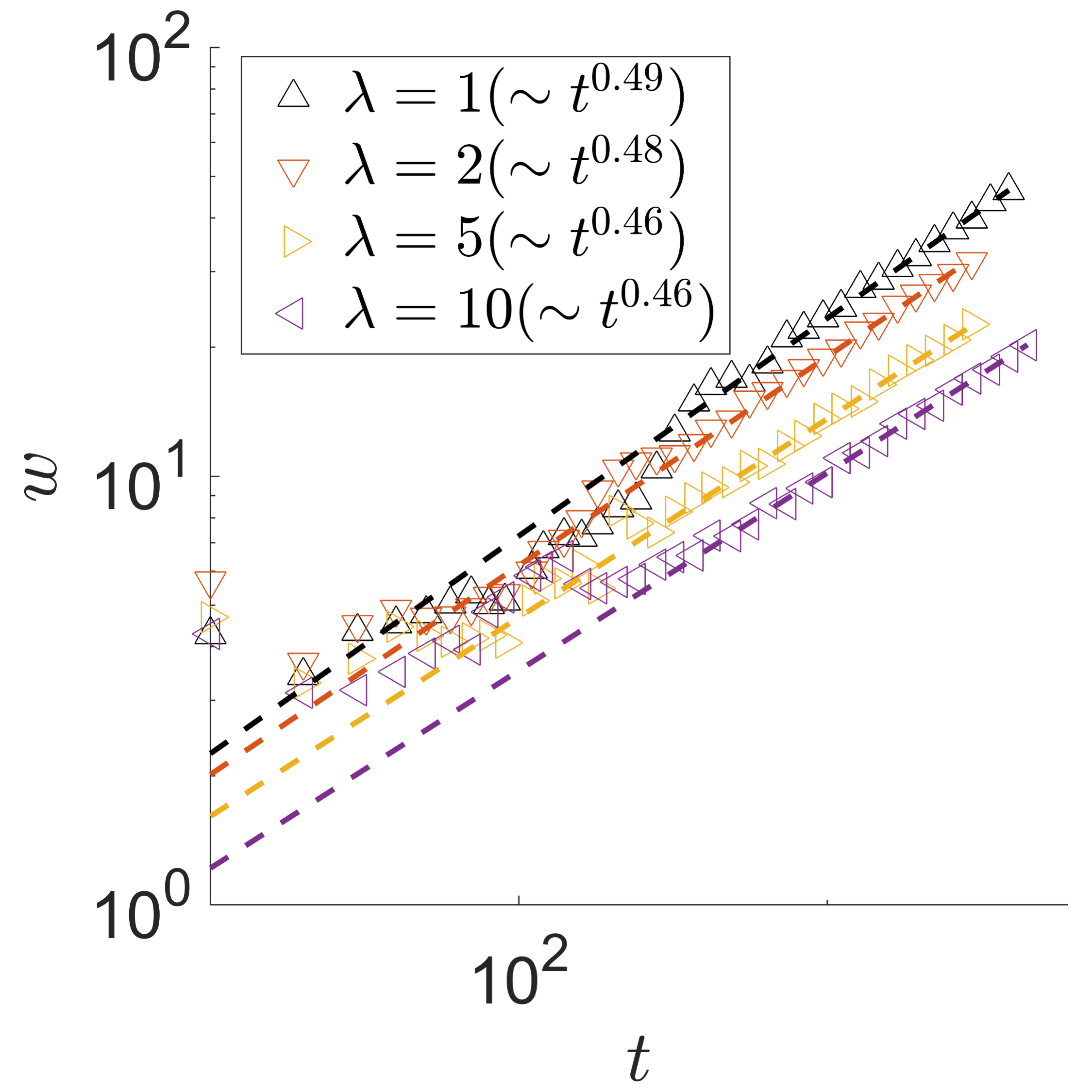}
\caption{{\bf Spiral branch width $w(t)$ of a spiral pattern.} $\lambda_{1,2} = 1, 2, 5, 10$. In all cases, $D_{A,B} = 1, \gamma_{A,B}=1, Y_{A,B} = 1$. The power-law fit is performed to times after a clear spiral is established.}
\label{fig:GR_spiral_growth_width}
\end{figure}

Moreover, the width of the growing branches show interesting dynamics. In our simulation, we measure the spiral branch width at a given time in the following way: We take the stretch of the branch still in contact with the open space, calculate the distance between the inner (in contact with the existing colony) and the outer (in contact with the open space) parts, and average over the entire stretch. As the colony continues to evolve, we see that $w(t) \sim t^{1/2}$ for $\lambda=1$ in Fig.\ref{fig:GR_spiral_growth_width}. The dynamics of the branch width $w(t)$ also shows a slight growth rate dependence.

In addition to characterizing the growing colony in terms of the colony radius $r(t)$ and wrapping branch width $w(t)$, we examine the angular position of the advancing branch tips. Setting a reference time $t_0$ when a clear spiral pattern emerges, we record the angular position of the advancing tips as $\theta_i(t_0)$, $i=1, 2$ in the case with 2 advancing spiral tips. We then measure the subsequent angular positions $\theta_{i}(t)$ with respect to $\theta_{i}(t_0)$. For example, after a transient of multiple interfaces merging around $t=250$ in Fig.\ref{fig:colony_spiral}, we set the angular position of the tips of the branches to be the reference angular position. 

\begin{figure}[!h]
\includegraphics[width=.5\textwidth]{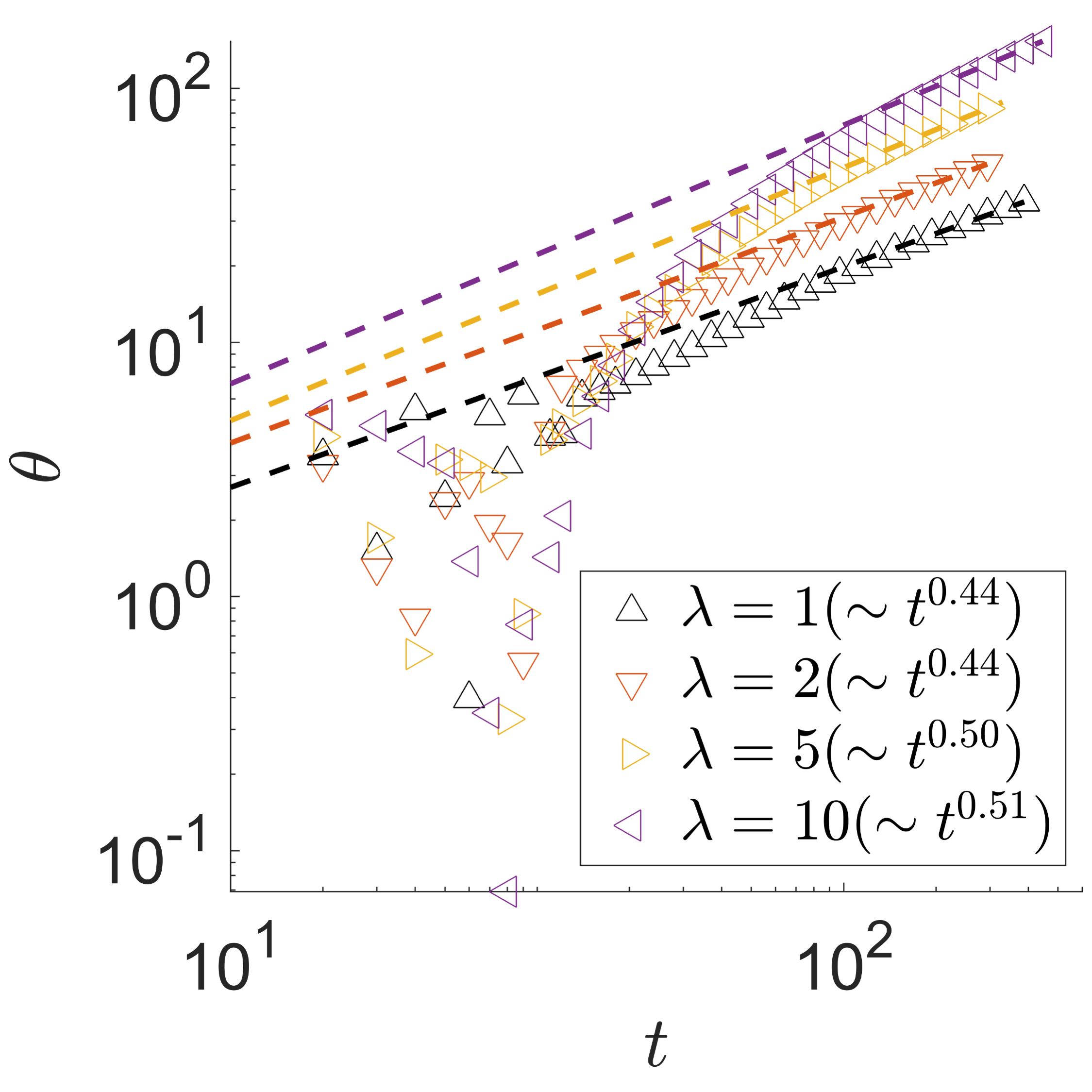}
    \caption{{\bf Tip angular position $\theta(t)$ of a spiral pattern.} $\lambda_{1,2} = 1,2,5,10$. In all cases, $D_{A,B} = 1, \gamma_{A,B}=1, Y_{A,B} = 1.$ $\theta(t=0)=0$ is chosen when the first angular position measurement is taken. The power law fit is performed between $t=10^3$ and $5\times t^3$.}
\label{fig:angle}
\end{figure}

In a stable spiral, we observe at least two advancing branches, one of each species. It is possible to have more as we will see later. This means both species can stably co-exist. We also note that as interfaces stochastically coalesce, the remaining growing tips of the two species settle into a stable angular separation around $\pi$, the maximum value in the case where there are two growing branches. At first, the growing tips are at random angular positions of the colony, depending on where the interfaces coalesce. When $t=250$ in Fig.\ref{fig:colony_spiral}, for instance, the angular separation between the two species' advancing front is between $\pi/2$ and $\pi$. In this case,
the consumer farther behind the producer (green species shown in Fig.\ref{fig:spiral_nutrient}) temporarily advance faster until the phase difference between the two advancing tips reaches the maximal value of $\pi$. The phase separation between the two advancing fronts then remains at $\pi$ given the symmetric choices of the parameters. The long-time dynamics of the tip angular position $\theta(t)$ also shows a slight growth rate dependence, as shown in Fig.\ref{fig:angle}. We show only one advancing front because the behavior of the other one is the same.

The branch width $w(t)$, the advancing tip angular position $\theta(t)$ and the colony radius $r(t)$ are inter-connected: In the steady state of two growing spirals, the \added{number of} newborn cells in a given time interval $dt$ equal to:
\begin{equation}
d\rho \approx 2 \times \left(w(t)\cdot r(t) \cdot d\theta+ \pi r(t) \cdot dw\right)
\end{equation}

The first term is the growth in the advancing tip and the second refers to the radial growth along the wrapping branch. As discussed before, the former contributes most of the colony growth. Our simulation results shown in Figs.\ref{fig:radius}, \ref{fig:GR_spiral_growth_width}, and \ref{fig:angle} give self-consistent scaling behaviors across the colony characteristics.

\begin{figure}[!h]
\includegraphics[width=.5\textwidth]{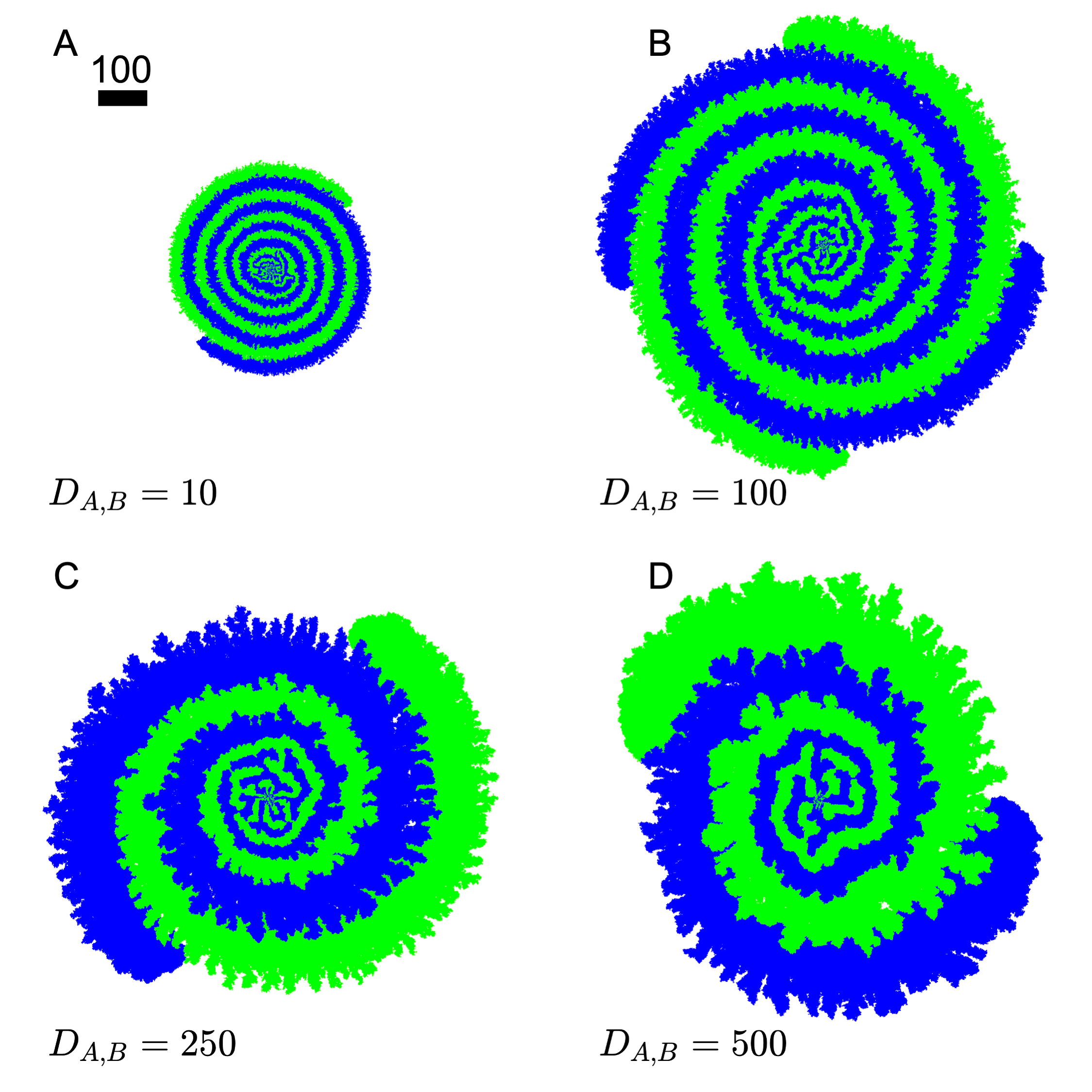}
\caption{{\bf Emergence of \deleted{island}\added{branch} structure with higher nutrient diffusion.} Colony snapshots at $t=400$ for different nutrient diffusion $D_{A,B} = 10, 100, 250, 500$. In all cases, $\lambda_{1,2} = 10, \gamma_{A,B} = 1, Y_{A,B} =1$}   \label{fig:island}
\end{figure}

The above discussions illustrate the spiral morphology when growth is limited by metabolite diffusion. It is natural to also look into the case when $D_{A,B}$ is increased for a specific growth rate. This time, we are surprised to see distinct \deleted{island-like}\added{branch} patterns emerging for very large $D_{A,B}$. Using $\lambda_{A,B} = 10$ for faster simulation results, we increased $D_{A,B}$ from 10 to 500, shown in Fig.\ref{fig:island}. As $D_{A,B}$ increases, the characteristic width of the branches increases as discussed above. Additionally, the outer edge of the colony develops individual  \deleted{``islands''}\added{branches}, reminiscent of diffusion limited aggregations seen in other computational models, for example in Refs. \cite{nadell2010emergence,tronnolone2018diffusion}. With higher nutrient diffusion, the characteristic width of the region where the cross-feeding metabolites exist is greater than that of the average width of the consumers. Therefore stochastically some consumer cells grow more radially than others, giving rise to the island structure. It is also worth noting that the spiral colony could have more than 2 growing branches, see for example the $D_{A,B} = 100$ case in Fig.\ref{fig:island} where 4 branches are present. And again, the phase difference between neighboring advancing tips is maximized at $2\pi/4$. Our simulations indicate that the number of branches is a stochastic result set in during the early stage of the colony when interfaces merge. For symmetry reasons, there is always an even number of branches, with two-branch spiral the most likely scenario.

\section{Summary and Discussions\label{sec:sum}}

Revealing and understanding the mechanisms of multi-species co-existence is a central question in ecological pursuits. The species coexistence emerges from spatial structure of their living community. Motivated by a theoretical cross-feeding model, our observations presented in this study suggest that the varying spatial availability of the cross-feeding metabolites can lead to drastically different, stable, and stunning, co-existence colony patterns.

Adopting a cellular automaton model on a square lattice using kinetic Monte Carlo simulation algorithm, we set out to study the spatial features of a two-species mutualistic cross-feeding mechanism. With symmetric parameters for both species, we discover non-trivial colony patterns with stable co-existence of both species as a result of cross-feeding nutrient availability through production and diffusion \added{for a wide range of individual cell growth rates and nutrient diffusion rates. We focus on the effects of $\lambda_{1,2}$ because they set the intrinsic timescale of the system. Together with the underlying lattice spacing and nutrient diffusion $D_{A,B}$, they also set the length scale of the system.} Qualitatively, the colony morphology evolves from a stable co-existing expanding sector\deleted{s} when nutrients are plentiful to a stable co-existing spiral with species grow along the interface. In the former regime, neither species need to rely on spatial proximity to the other. The initially mixed inoculum self-separates and grows into radially expanding sectors separated by super-diffusive interfaces. In the latter regime, nutrient availability is localized near the producer cells. Within the growing colony, instead of radially expanding, the interface between the two species bends and each species grow\added{s} along the other. Transitioning between the radially extending interfaces and the spiral interfaces, there is also the possibility of interfaces coalesce, resulting in one species being engulfed by the other. Beyond analyzing the macroscopic colony patterns, we also investigated the colony characteristics including colony radius, and the spiral branch width.

\added{Spiral patterns and spiral waves are commonplace in nature, occurring both in biological systems and chemical reactions. What we revealed here is a general mechanism for spiral patterns to emerge in a generic cross-feeding two-species system.} The lattice-based model is versatile and readily adaptable to explore the vast parameter space outlined in this work. Here we focus on the symmetric mutualistic interaction between the two species. \added{We have also performed preliminary studies on cases with asymmetric growth rates and diffusion coefficients, discussed in S2. We see the preservation of the spiral pattern with small differences between either the growth rates, or the diffusion coefficients. This promises a wider parameter space where we can observe spiral patterns as a result of mutualistic cross-feeding. Beyond our study presented here, we} 
\deleted{We}see multiple promising directions for further investigation: On the theoretical front, we can adapt the system to studying other types of cross-feeding such as commensalism and syntrophy. It is also possible to incorporate negative effects of the accumulation of the excess metabolites, as suggested by recent experiments \cite{goldschmidt2018metabolite}, that impact local species diversity. \added{At present, the metabolic rule is such that for cells that do not divide due to limitation of space, they continue to take up and excrete metabolites as a form of cell maintenance. This condition can be revised for different growth limitations and to incorporate cell deaths.} The simple construction of this model also serves as an ideal primer to quickly scan the parameter space before extending into more sophisticated, yet computationally more intense, three-dimensional models such as Ref. \cite{warren2019spatiotemporal}. In this study, we focus on the diffusion of the cross-feeding metabolites. We will direct our future work in including cell motility \added{inter-cell mechanical interactions} to see how that impacts spatial structures of the colony.

\section*{Supporting information}
\paragraph*{S1 \added{Measurement of interface fluctuations.}\label{S1}}
{\bf Measurement of interface fluctuations.} We develop the algorithm in MATLAB code. To characterize the fluctuating boundaries separating the two species, we define the lattice site $p_i$ along the interface as the following: Each lattice site can take on values 0 (empty), or 1 or 2 depending on the occupant cell type. $p_i$ is {\it on the interface} if it is occupied by a cell and there are only two unique non-zero values among itself and its four nearest-neighbor sites. The collection of points $\{p_i\}$ defines the interface  and its length is $L$.

For each interface, we quantify the fluctuation along the interface as a function of a sliding window $x \in [1, L]$. Given $x$, We first find the best linear fit $y(x')$ for the segment connecting from $p_0$ to $p_{x-1}$. We then compute the mean square displacement $y_0^2$ between the interface and $y(x')$. This process is repeated by moving the segment along the interface one point at a time. Then the average of $y_i^2$ gives us the mean square displacement $\overline{y^2}(x)$ for a given window size $x$.

\paragraph*{\added{S2 Colony patterns with asymmetric mutualism.}\label{S2}}
\label{app:asym}
{\bf Colony patterns with asymmetric mutualism.} Our work focuses on the mutualistic cross-feeding case where all parameters are the same for both species. However, the spiral patterns are robust even when the parameters are asymmetric. Below we show a few scenarios where the nutrient diffusion rates and the growth rates are different for the two species. Given the large parameter space, we will reserve a comprehensive study on all possible asymmetric configurations in another study.

\begin{figure}
\centering
\includegraphics[scale=.75]{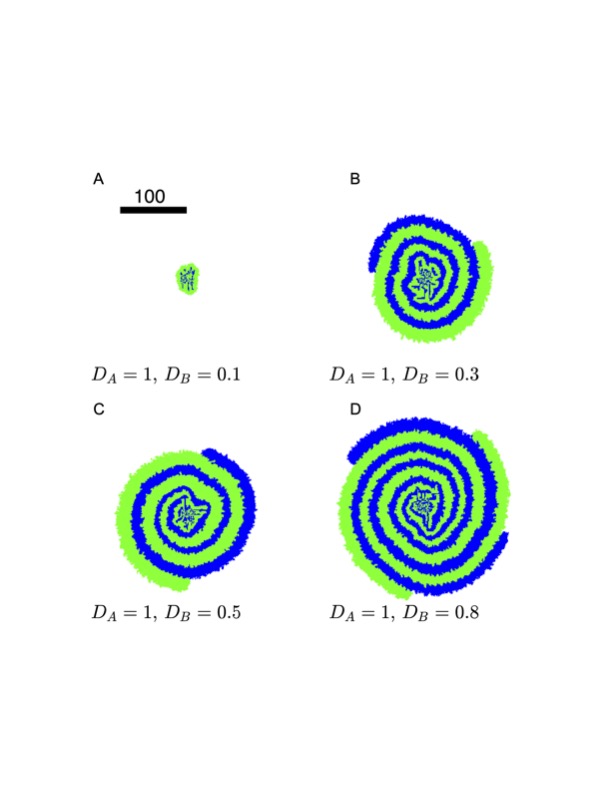}
\caption{{\bf Colony patterns with asymmetric nutrient diffusion rates.}
Colony snapshots at $t=1000$. Equal number of species 1 and 2 are seeded in a $20 \times 20$ patch with initial density $\rho_0=1/4, \lambda_{1,2} = 1$, $\gamma_{A,B} =1$, $Y_{A,B}=1$, $D_A= 1$ and $D_B$ = A) 0.1, B) 0.3, C) 0.5, and D) 0.8. Scale bar indicates the width of 100 cells. Green cells are species 1, and blue ones are species 2.}
   \label{fig:D}
\end{figure}

As one of the nutrient diffusion rate, $D_B$ (molecules excreted by species 1 to be taken up by species 2), is reduced while keeping $D_A$ the same, we see the system continue to emerge into a stable spiral pattern, shown in Fig.\ref{fig:D}. When $D_B \ll D_A$ with the same excretion rate, the slower-diffusing nutrient leads to local concentration lower than the Monod constant $K_B$, thus reducing the effective growth rate for species 2 (blue cells). In this case, species 2 lose out in competing for available growth space and is engulfed by the faster growing species 1, as shown in Fig.\ref{fig:D}(A). The decrease in one of the nutrient diffusion rates also led to the decrease in overall colony size due to the coupling of individual growth rates $\lambda_{1,2}$ and local nutrient concentration $n_{A,B}$.

\begin{figure}
\centering
\includegraphics[scale=.75]{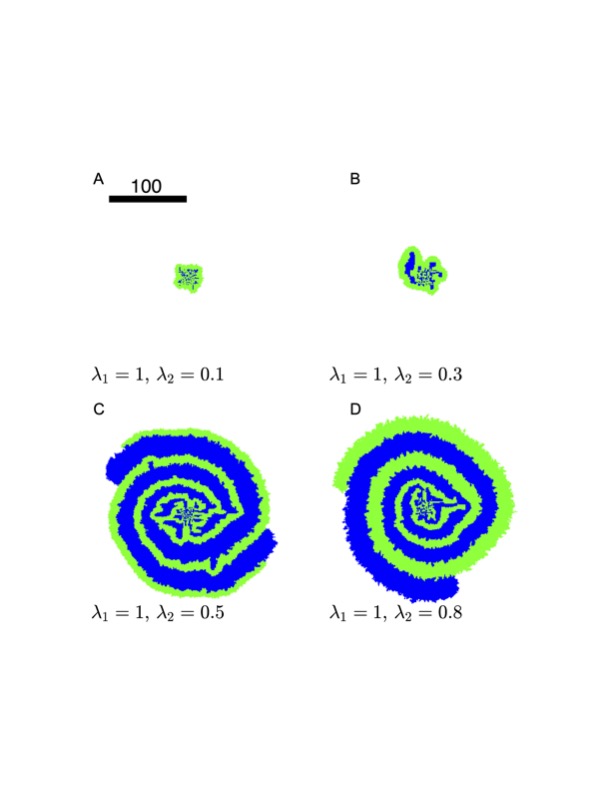}
\caption{{\bf Colony patterns with asymmetric cell growth rates.}
Colony snapshots at $t=1000$. Equal number of species 1 and 2 are seeded in a $20 \times 20$ patch with initial density $\rho_0=1/4$, $D_{A,B}=1$, $\gamma_{A,B} =1$, $Y_{A,B}=1$, $\lambda_A=1$, and $\lambda_B$ = A) 0.1, B) 0.3, C) 0.5 and D) 0.8. Scale bar indicates the width of 100 cells. Blue cells are species 1, and green ones are species 2.}
   \label{fig:GR}
\end{figure}

When the maximal individual cell growth rate $\lambda_{1,2}$ is varied, we again observe a stable spiral pattern when $\lambda_1$ and $\lambda_2$ are comparable. In Fig.\ref{fig:GR}(A-B), species 2 (blue cells) have a much smaller maximal growth rate. At the nascent stage of the colony, the limited number of empty lattice sites are taken up by species 1 (green cells), resulting in the engulfment pattern. When $\lambda_2$ is increased, we see the emergence of a stable spiral pattern again as shown in Fig.\ref{fig:GR}(C-D). It is  worth noting that when the two species have different growth rates, the slower growing cells (blue in Fig.\ref{fig:GR}) develop a wider wrapping branch, which is consistent with what we discussed in the main article.

\paragraph*{S3 \added{All codes used to generate the data are published on\href{https://github.com/mr7Jacky/mutualistic-crossfeeding}{Github}}\label{S1}}
All codes used to generate the data are published on 365 Github: https://github.com/mr7Jacky/mutualistic-crossfeeding
\label{app:interface}

\section*{Acknowledgments}
The authors acknowledge the financial support from the National Science Foundation through grants DMR-1702321, MCB-2029480, and MCB-2029580. This work benefited from fruitful discussions with T. Hwa and B. Li. JJD is grateful for the hospitality of T. Hwa at UCSD during her sabbatical visit, where this project initiated.

\bibliographystyle{plos2015.bst}

\end{document}